%% file: ms.tex
\documentclass[12pt,preprint]{aastex}
\shorttitle{XRT/Optical Positions}
\shortauthors{Butler}

\def\gtrsim{\mathrel{\hbox{\rlap{\hbox{\lower4pt\hbox{$\sim$}}}\hbox{$>$}}}}
\def\lessim{\mathrel{\hbox{\rlap{\hbox{\lower4pt\hbox{$\sim$}}}\hbox{$<$}}}}
\newcommand{\beq}{\begin{equation}}
\newcommand{\eeq}{\end{equation}}

\begin{document}

\title{Refined Astrometry and Positions for 179 Swift X-ray Afterglows}

\author{N.~R. Butler\altaffilmark{1,2}}
\altaffiltext{1}{Townes Fellow, Space Sciences Laboratory, University of 
California, Berkeley, CA, 94720-7450}
\altaffiltext{2}{Astronomy Department, University of California,
445 Campbell Hall, Berkeley, CA 94720-3411, USA}

\begin{abstract}
We present a refined catalog for the positions of 179 $\gamma$-ray burst (GRB)
X-ray afterglows observed by the X-ray Telescope (XRT) on Swift prior to November 1, 2006.
The positions are determined by detecting
X-ray field sources in the deep X-ray images and comparing the
centroids to those of optical sources in the Digitized Sky
Survey (DSS) red2 catalog or the Sloan Digital Sky Survey (SDSS) DR-5 catalog.
Half of the 90\% confidence error region radii are $<2.2$ arcseconds.
The error regions areas are typically $\sim 4$ times smaller than the best XRT-team error
regions, although the positions require deep X-ray integration ($>20$ ksec)
and cannot be generated nearly as rapidly after the GRB.
The positions derived for $>90$\% of 77 bursts with optical afterglows are
consistent with the optical transient positions, without the need for
systematic error.  
About 20\% of the afterglows positions require a sizable shift in the Swift
satellite aspect.  
We discuss the optical/X-ray properties of the field sources
and discuss implications of the frame offsets for studies of optically dark
GRBs.
\end{abstract}



\section{Introduction}

The {\it Swift}~satellite \citep{gehr04} is revolutionizing our understanding of $\gamma$-ray
bursts and their afterglow emission.  Key to {\it Swift}'s success are the rapid
localization capability and unprecedented detection sensitivity of the Burst
Alert Telescope \citep[BAT;][]{bart05} and the capability of the X-ray Telescope 
\citep[XRT;][]{bur05} to determine
$\sim 5$\arcsec~positions for the burst afterglows within minutes of the BAT
$\sim 3$\arcmin~source localizations.  These localizations have enabled the
detection and study of $\sim 80$ GRB optical afterglows since the satellite launch
in late 2004 and prior to November 1, 2006.

The pointing direction of the {\it Swift}~satellite is determined by star trackers with an
accuracy $\lessim 5$\arcsec.  However, X-ray sources are typically centroided much more finely ($\lessim 1$\arcsec),
and the GRB afterglow position errors quoted by the XRT team are dominated by systematic error terms which
account for the poor relative accuracy of the star trackers.
Recently, \citet{mor06} have refined the XRT boresight calibration (which relates the
XRT alignment to the star tracker positions) and effectively
reduced the systematic position error by a factor of two for 68 afterglows.  Here, we present 
a methodology for sidestepping the satellite instrument alignment and pointing uncertainties
which decreases the position errors by an additional factor of two.
We present refined X-ray positions for 179 bursts determined by matching X-ray field source positions
directly to those of counterpart optical sources.  The sample of 179 
bursts is that for which the XRT team has published positions in the $\gamma$-ray Coordinates
Network (GCN) circulars and in \citet{mor06}.

\section{X-ray Source Detection and Centroiding}

The {\it Swift}~XRT Position Counts (PC) mode data are reduced by running
the {\tt xrtpipeline} reduction task from the
HEAsoft~6.0.6\footnote{http://heasarc.gsfc.nasa.gov/docs/software/lheasoft/}
software release and using calibration files from the most recent
calibration database release (2006-04-27). 
Starting with a position guess for the afterglow, we map the counts
from each followup observation onto a tangent plane centered at that
position.  Detections of and positions for X-ray field sources in the composite image
with signal-to-noise ratio $>3$ are found by running {\tt wavdetect} \citep{fre02}
on scales $2,2\sqrt{2},4,4\sqrt{2}$,and 8 pixels.  
To additionally reject spurious sources, we require the {\tt wavdetect} sources to have
an estimated size $>$5 pixels and to have aspect ratios $<2$.
We also reject highly variable X-ray (KS-test prob. $<10^{-10}$) sources, 
which are typically
spurious sources or sources off-chip for a portion of the observation.

Special attention is paid to the centroiding of the afterglow, which is taken
to be the most significant and variable source within a $50\times50$ pixel$^2$
region about the position guess.  The XRT team positions are used for the position
guess.  In the few cases where the afterglow faintness leads to more than one
possible candidate, we adopt the candidate selected by the XRT team.
A maximum likelihood (ML) centroiding algorithm is used for the XRT PC
mode counts extracted in a $20\times20$ pixel$^2$ region about this source.

The likelihood for the observed counts $i$ (positions $x_i$, $y_i$ and energy $E_i$)
as a function of the source position ($x_{\circ}$,$y_{\circ}$) can be written:
\beq
{\cal L} = C^N~ \left \{ \prod_{i=1}^{N} \Phi_{\rm psf} (x_i-x_{\circ},y_i-y_{\circ},E_i) \right \} 
\exp \left \{ -C\int \!\!\! \int dxdy ~\Phi_{\rm psf}(x-x_{\circ},y-y_{\circ})~ M(x,y) \right \},
\eeq
where $C$ is a normalization, $M$ is a (source-flux-weighted) exposure map,
and $\Phi_{\rm psf}$ is the point spread function model from the calibration database 
({\tt swxpsf20010101v003.fits}).
The
term in the exponential rigorously accounts for the presence of bad or missing 
detector columns or pixels.  We maximize this expression with respect to
$C$, and use the following fit statistic for $N$ total counts for the centroiding:
\beq
 \chi^2 = 2N\log\left \{\int \!\!\! \int dxdy ~\Phi_{\rm psf}(x-x_{\circ},y-y_{\circ})~ M(x,y)
\right \} - 2\sum_i \log [ \Phi_{\rm psf}(x_i-x_{\circ},y_i-y_{\circ},E_i) ].
\eeq
Circular error regions are determined by numerical integration for each
centroid.

The calculated position error at 90\% confidence versus the number of detected
source counts $C$ for 179 burst afterglows is well described as a powerlaw
$13.4 ~C^{-1/2}$ arcsec, with significant scatter below 20 counts.  
Comparing the centroids for the bursts with optical transient (OT) positions, 
we find that the ML centroids are on average 10\% more accurate than centroids 
determined using the {\tt xrtcentroid} task.  The empirical errors from that 
tool and XRT data have been found to follow $23 ~C^{-0.48}$ \citep{hill05,mor06},
which is typically a factor $\sim$2 larger 
than for the ML centroider.

\section{Probabilistic X-ray Image Registration to Optical Catalogs}
\label{sec:match}

We begin with a deep X-ray exposure and the comparison of the X-ray source positions to the 
positions of optical sources.
If data from the Sloan Digital Sky Survey \citep[SDSS;][]{am06} are present
for a given X-ray afterglow field, we download the catalog data and
fits images.  We also download the Digitized
Sky Survey (DSS2-red) images and the USNO-B2 catalog for the field.
We then run {\tt sextractor} \citep{bertin96} over the images to generate the deepest
possible catalog ($r'\lessim 22.5$ for SDSS, $R\lessim 20$ for DSS).  
The DSS astrometry is corrected to sub-arcsecond levels using the Two Micron All 
Sky Survey.

\begin{figure}
\includegraphics[width=3.7in]{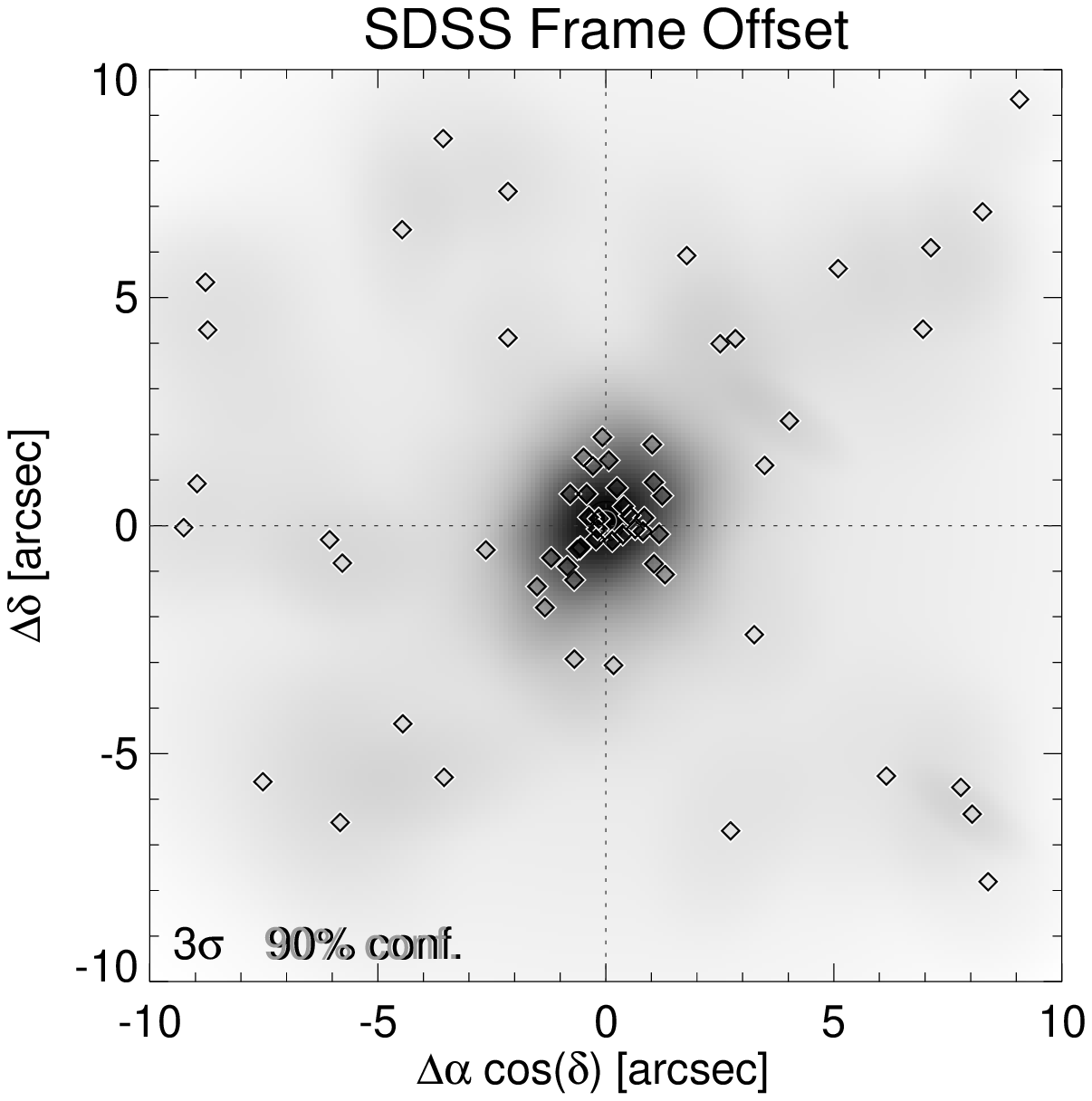}
\includegraphics[width=3.7in]{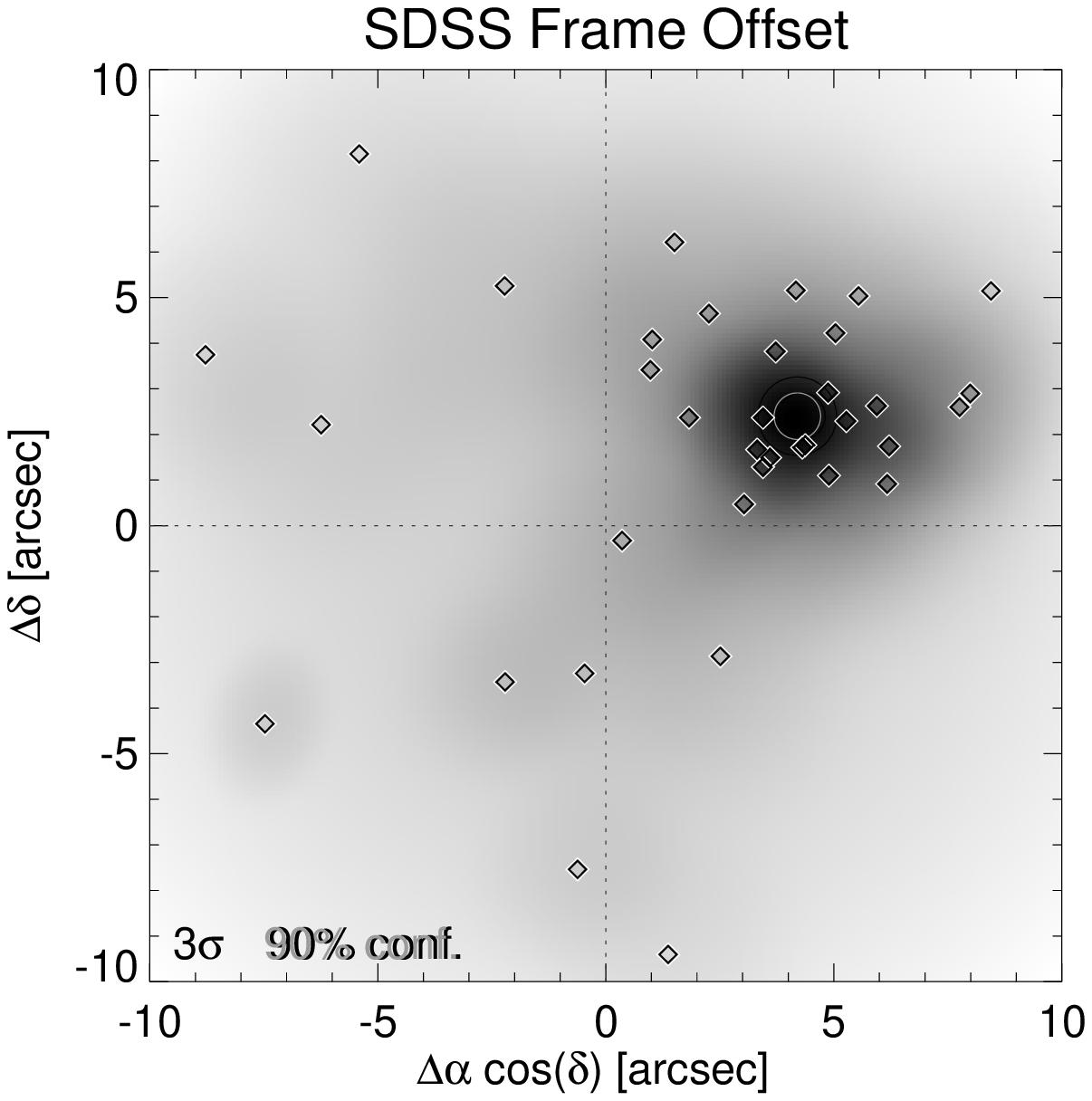}
\caption{\small Example frame offset probability plots for GRB~050408 and GRB~060108.
The points are the SDSS optical source positions relative to the X-ray source
positions.  The shaded background describes the offset probability between
the optical and X-ray field sources (Section 
\ref{sec:match}).  For GRB~050408, the best fit frame offset is
$\Delta \alpha= 0.0$, $\Delta \delta= -0.1$, $\pm 0.3$ [arcsec] 90\% Conf.
For GRB~060108 \citep{bnb06,oates06}, the best fit frame offset is
$\Delta \alpha= 4.2$, $\Delta \delta= 2.4$, $\pm 0.5$ [arcsec] 90\% Conf. 
We also plot 90\% confidence and $3\sigma$ error circles.}
\label{fig:f1n2}
\end{figure}

For each {\tt wavdetect} X-ray source $i$, we record the positions of optical sources
$j$ within a box of area $A$ equal to $20\times20$ arcsec$^2$.  We consider each optical
source an equally likely possible counterpart to X-ray source $i$.  The 
probability distribution for the offset of the X-ray image from the optical 
image, given this one source, can be written:
\beq
P_{{\rm X},i} = {P_0 \over A} + (1-P_0)~ \sum_j ~G( [x_j-x_i]^2+[y_j-y_i]^2 , \sigma_i ),
\eeq
where the sum of Gaussian terms $G$ is normalized over the allowed area $A$.  
Here, the positions $x$ and $y$ are related to shifts in right ascension $\Delta \alpha$ 
and declination $\Delta \delta$ approximately by $x=\Delta\alpha \cos(\delta)$ and $y=\Delta \delta$,
both in arcseconds.
The prior-probability of finding an optical/X-ray match is $P_0$.  Based on fields searched prior
to 2006 (see also Section \ref{sec:match_prop} below), we set these to be
$1/2$ and $1/3$ for the SDSS and DSS, respectively.  We allow for the
possibility of a small frame rotation be replacing each term $G$ in equation (3) with 
\beq
\bar G = {1 \over \sqrt{2\pi\sigma_{\theta}^2} }\int d\theta ~G( [x_j-x_i-\theta (y_i-y_p)]^2,[y_j-y_i+\theta (x_i-x_p)]^2, \sigma_i ) ~\exp{\left \{-{\theta^2 \over 2\sigma_{\theta}^2}\right \}},
\eeq
where $(x_p,y_p)$ is the XRT aim point.  In marginalizing over the unknown rotation angle, we assume a Gaussian prior
with $\sigma_{\theta}=5$\arcmin.
Combining the information from each X-ray field source, the offset probability can be written:
\beq
 P_{\rm X-opt} = {1 \over 2\pi \sigma_r^2} \exp{\left \{ -{x^2+y^2 \over 2\sigma_r^2 } \right \} } ~ \prod_i P_{{\rm X},i},
\eeq
where we use a Gaussian prior that the offset radius is $0\pm \sigma_r$, with $\sigma_r=3$\arcsec.  
Figure \ref{fig:f1n2}
shows example Optical---X-ray positions and probabilities derived for the GRB~050408 and GRB~060108 fields.  In the backgrounds, we plot
$-2\log ( P_{\rm X-opt} )$.

As we discuss below, a relatively long total exposure ($>20$ ksec) over several
orbits is required to detect and match X-ray field sources.
Because the satellite re-acquires the field every orbit, we expect (and observe)
there to be a scatter in the aspect solution for different visibility periods.
It is therefore possible that the position offset for the period of bright GRB afterglow activity
is not the same as the mean frame offset determined from the full integration.

To account for the effect of shifting aspect during the observations of an
afterglow over several days or weeks, we co-add the counts in $20\times 20$ pixel$^2$
regions about each X-ray field source.  An additional composite field star image is
formed by replacing the total field star counts in each good time interval of
data acquisition by the counts detected for the afterglow in that interval.
We then determine the afterglow offset relative to the mean frame offset 
by spatially shifting the counts-weighted composite field star image until it
overlaps with the first composite field star image.  This calculation typically contributes error
$\sim 1$ arcsec, although this can be considerably larger for faint or
rapidly fading sources.

\section{Discussion}

\subsection{Comparison With Optical Transient and XRT Team Positions}

\begin{figure}
\centerline{\includegraphics[width=3.7in]{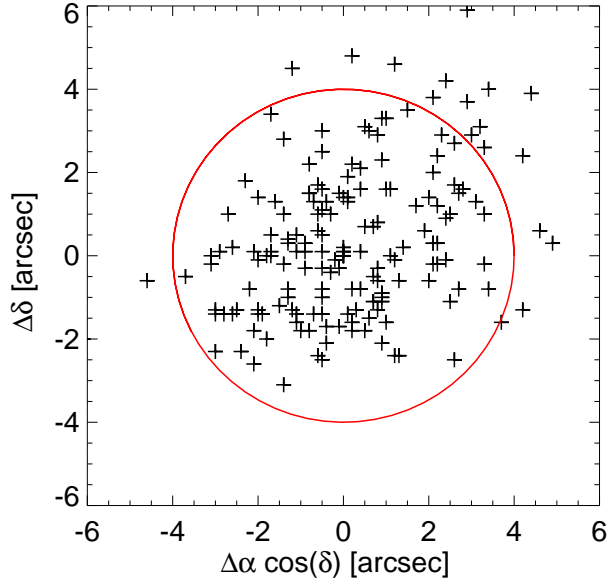}}
\caption{\small
XRT Frame Offsets for 179 Bursts relative to DSS/SDSS.  90\% of the offsets are contained
within the red, 4\arcsec~radius circle.  This radius is comparable to the XRT team systematic
error.}
\label{fig:f0}
\end{figure}

\begin{figure}
\centerline{\includegraphics[width=4in]{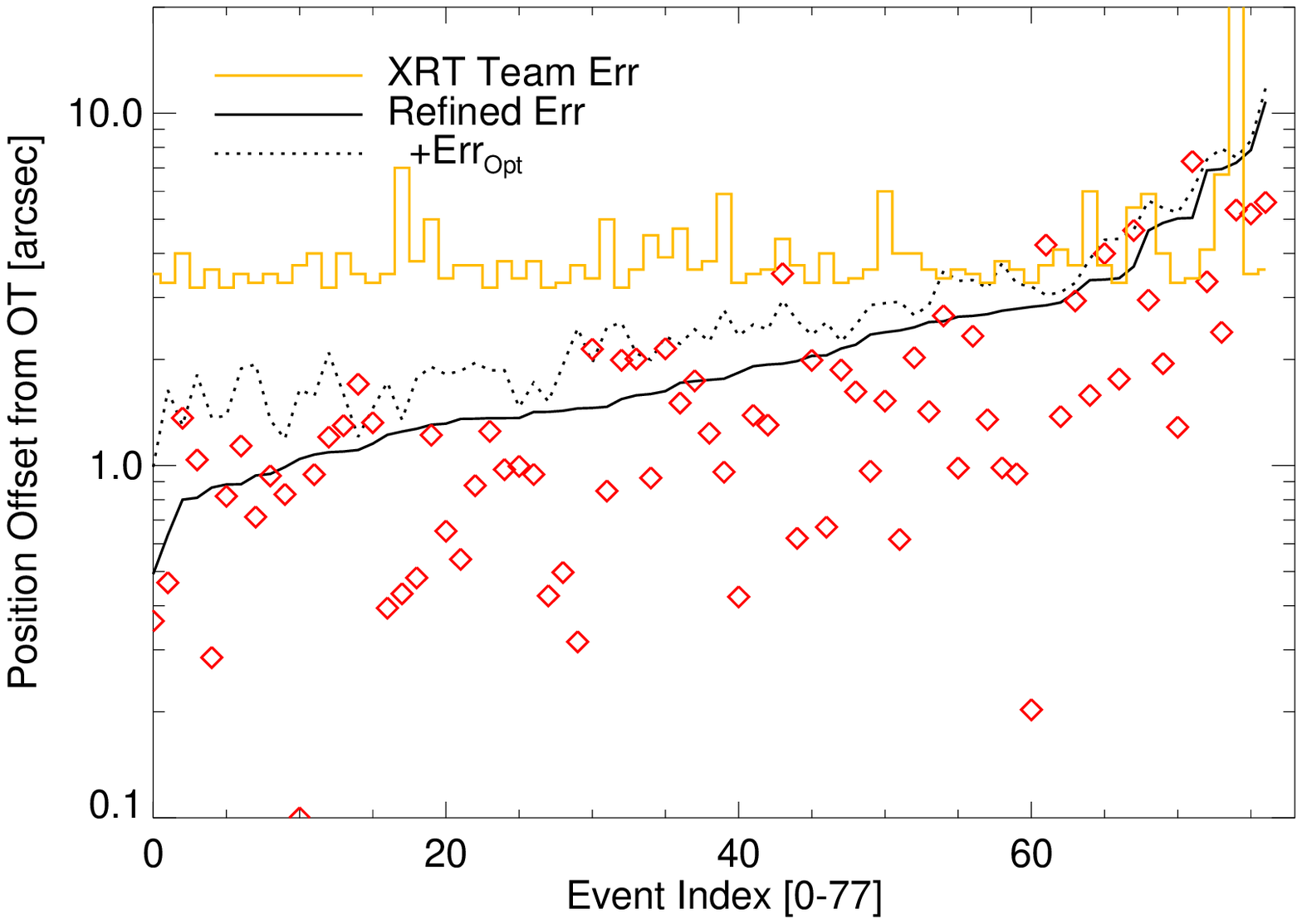}}
\centerline{\includegraphics[width=4in]{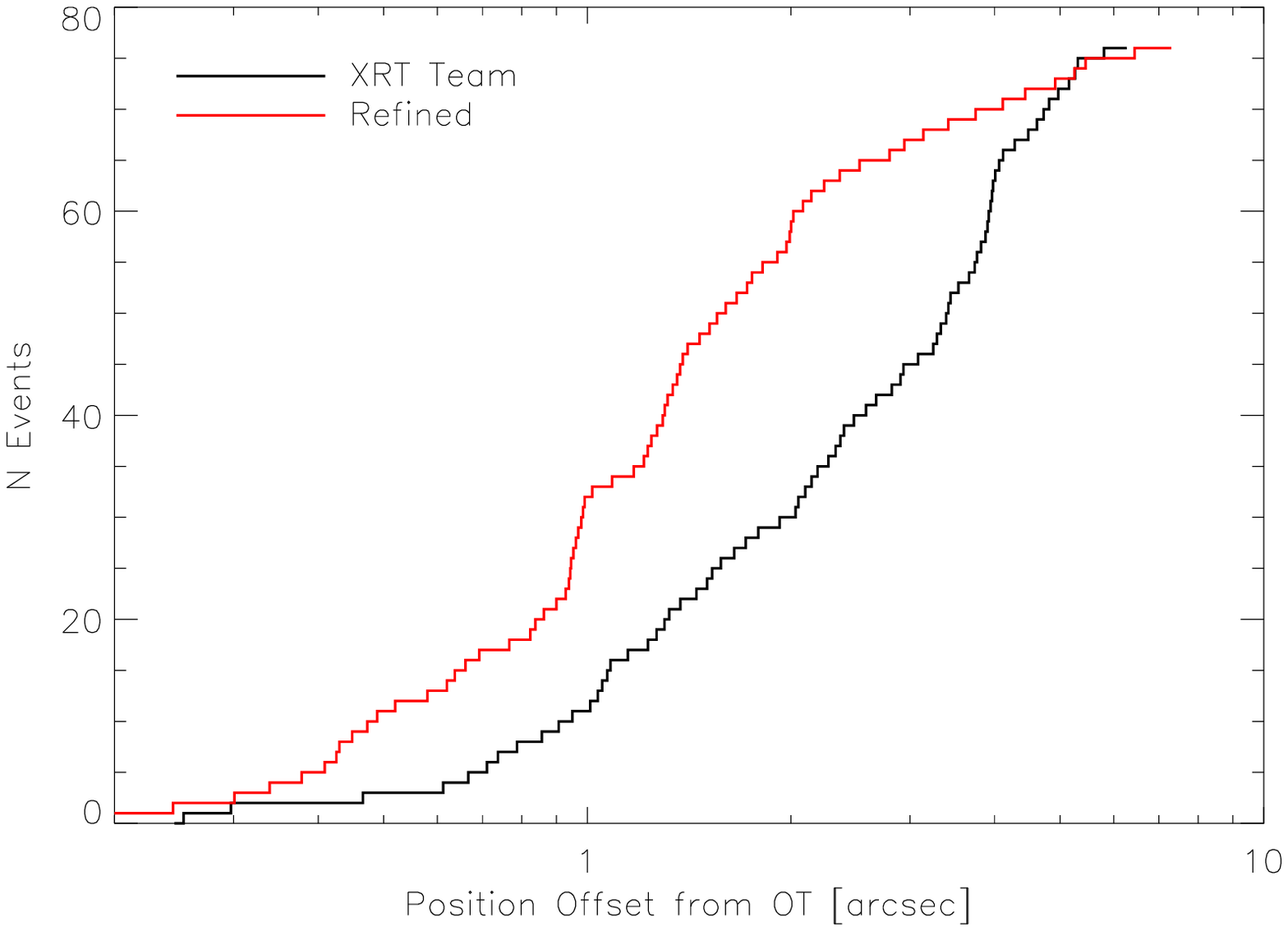}}
\caption{\small (A) Comparison of the final X-ray position differences from the
optical transient positions in the literature (red points).  The X-ray position
90\% confidence error is overplotted (solid black curve), as is the X-ray plus OT
position error (dotted curve).  The median offset is 1.3 arcsec.  The median
error (X-ray only) is 1.8 arcsec.  Of the X-ray positions, $71/77=92$\% are consistent with the
optical positions.  The orange (histogram) curve shows the XRT team error radius for each event.
The offset of the XRT team positions from the OT positions are not plotted. However, we note
that $68/77=88$\% of the XRT team positions are consistent with the OT positions.
(B) Cumulative plot of the positions offsets relative to the OT positions for the refined
and XRT team positions.  The nearly constant logarithmic shift over a range of offsets corresponds
to a factor of two increase in positional accuracy.  From a KS-test, the probability that the two
offset distributions are drawn from the same distribution is $10^{-5}$.
}
\label{fig:f3}
\end{figure}

Tables 1---3 give our X-ray positions for 179 {\it Swift}~GRB afterglows with X-ray positions
quoted in the GCN.  In most cases, we cleanly detect the X-ray transient and observe it to fade
in time.
In a few cases (GRB~050906 GCN3956, GRB~050911 GCN3967, GRB~050925 GCN4043, GRB~060223B GCN4835) 
the fades are only weekly confirmed, consonant with the XRT Team's reports
that these sources are only candidates.   These positions should be treated with caution.
Figure \ref{fig:f0} shows the best-fit offset calculated for each of the 179 afterglows.
Ninety percent of the offsets are $\lessim 4\arcsec$.  The scatter is similar to the
systematic error (3.2\arcsec, 90\% conf.) introduced in \citet{mor06}.
Both the position offsets relative to the known OT positions (Figure \ref{fig:f3})
and the reported errors (Figure \ref{fig:f3}A) decrease by 50\% on average, relative
to those reported by the XRT team.
We find that $\gtrsim 90$\% of the 77 source positions lie within our computed error
(statistical only) of the OT positions, also accounting for the error in the OT positions.
A large fraction $\sim 90$\% of the XRT Team positions are also consistent with the OT positions.
Finally, we note that $40/179=22$\% of our positions lie beyond the quoted XRT error radius from
the XRT team position.  The positions for these bursts are underlined in Tables 1---3.  If
we restrict to just the bursts with refined positions quoted in \citet{mor06}, 13\%
of the positions are discrepant.

\subsection{Properties of the Optical/X-ray Field Sources}
\label{sec:match_prop}

In this and the next section, we consider {\it Swift}~bursts detected prior to September 11, 2006,
corresponding to the date of submission of this paper.  For this time period,
49\% (35\%) of the X-ray field sources have optical sources in the SDSS (DSS) within
2\arcsec.  This justifies our choice of priors on the match probability (Section \ref{sec:match}).
The choice of priors leads to tighter error bars from an SDSS catalog match than for the DSS catalog match.  

\begin{figure}
\centerline{\includegraphics[width=5.0in]{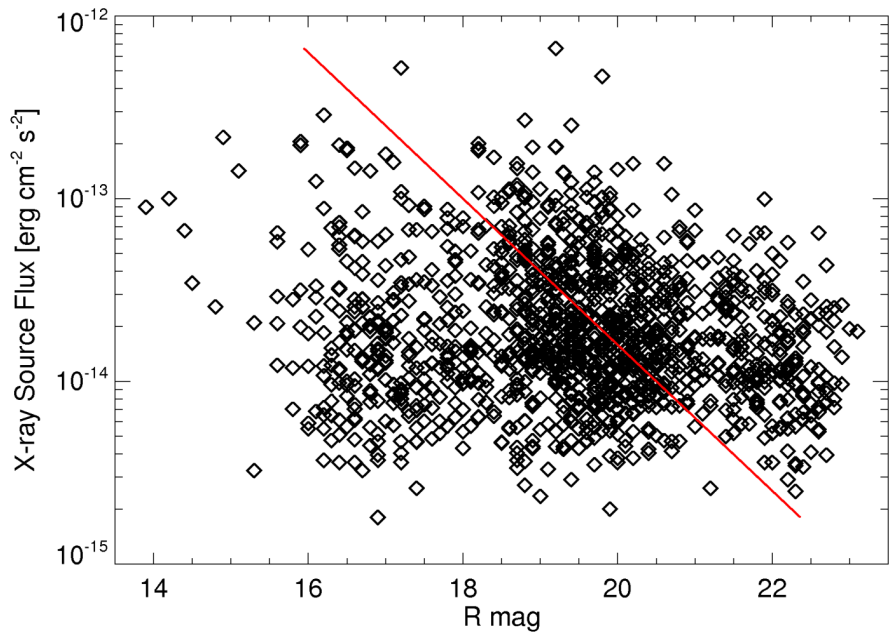}}
\caption{\small The X-ray fluxes and optical magnitudes for matched sources between the XRT and DSS/SDSS.
Missing from the plot are the non-detections ($R>22$) for $\sim 1/3$ of the X-ray sources.
The solid red line shows the best fit curve from deep field observations with {\it ROSAT}~and {\it Chandra}~\citep{has98,giac01}.  There
is considerable scatter about that line.}
\label{fig:f4}
\end{figure}

From deep field observation with {\it Chandra}~\citep{giac01}, $\approx 90$\% of X-ray sources brighter than $F_{\rm X}\approx
3\times 10^{-16}$ erg cm$^{-2}$ s$^{-1}$ have optical counterparts brighter than $R=26$.  
The {\it Chandra}~source counts at higher flux levels \citep[Figure 7][]{giac01} appear consistent with the detection fraction
quoted above, where $F_{\rm x} > 2 \times 10^{-15}$ erg cm$^{-2}$ s$^{-1}$ and $R<20$ (DSS) or $r'\lessim 22.5$ (SDSS).
There is considerable scatter about
the $F_{\rm X}/F_{\rm opt}=1$ line (solid line in Figure \ref{fig:f4}), which describes {\it ROSAT} sources in the Lockman Hole \citep{has98}.
We observe a weak $\tau_{\rm Kendall}=-0.09$ (although $5\sigma$ significant) correlation of the optical $R$-band magnitude with
X-ray flux.  There is a large scatter $\sim 1.7$ mag, roughly independent of the X-ray flux.  Here, we ignore the difference between
the $r'$ and $R$ bands, and we adopt a constant scaling of $5 \times 10^{-11}$ erg cm$^{-2}$ rate$^{-1}$ in order to convert the
observed XRT rate in counts per second to flux.  

\subsection{Implications for Optically Dark Bursts and GRB Host Galaxies}

Our refined positions for 39 of 160 X-ray afterglows lie outside of the error regions determined by
the XRT team.  This fraction is one possible metric on the fraction of potential missed optical source associations---
either afterglows or host galaxies---because even modest shifts in the refined positions relative to those quoted by the XRT team can
allow previously excluded regions of sky and can have important consequences for the optical source
associations \citep[see, e.g.,][]{bloom06}.  

The fraction of missed afterglows must be smaller than $39/160$, because bright afterglows should
be detected independent of modest X-ray positions misdeterminations.  An OT is most likely to be detected either
when the position lies inside the XRT team error region or when the OT is sufficiently bright.  As a proxy for the fraction of bright
OTs, we take the fraction of the 39 bursts with poor positional agreement, for which an OT was nonetheless detected ($P_{\rm bright}=14/39$).
Faint afterglows would have been detected for only the bursts with excellent positional agreement ($P_{\rm good~pos}=121/160$).
We should then find that $P_{\rm bright} + P_{\rm faint}~P_{\rm good~pos} = 69/160$, the observed OT detection
fraction.  The bursts are either optically bright, faint, or dark.  From this, we calculate the fraction of optically faint
bursts $P_{\rm faint}\approx 9$\%, implying that only $\approx 4$ OTs may have been missed in the sample of 39 X-ray afterglows with
position offsets.  One well-documented example of such a case is GRB~060108 for which
our SDSS-refined X-ray position \citep{bnb06} led to the detection of the infra-red afterglow \citep{oates06},
located just outside the XRT-team refined error region.  

Although we conclude that future refined positions are unlikely to dramatically increase the number of OTs detected,
we do plan to make the positions available to the community as rapidly as possible.
X-ray field source detection apparently works well for afterglows observed for $\Delta t \gtrsim 20$ ksec,
with the tightest possible error regions becoming possible for $\Delta t \gtrsim 50$ ksec.  These exposures
have typically occurred within 1 to several days after a bursts.  Our positions are autonomously determined 
on these timescales and are published on the web\footnote{http://astro.berkeley.edu/$\sim$nat/swift/xrt\_pos.html}.  Because the position and error radius change in 
time as more data are accumulated, consumers should quote the version number
(incremented from the version 0.0 positions quoted here) when referencing
one of these positions.

\begin{figure}
\centerline{\includegraphics[height=5in]{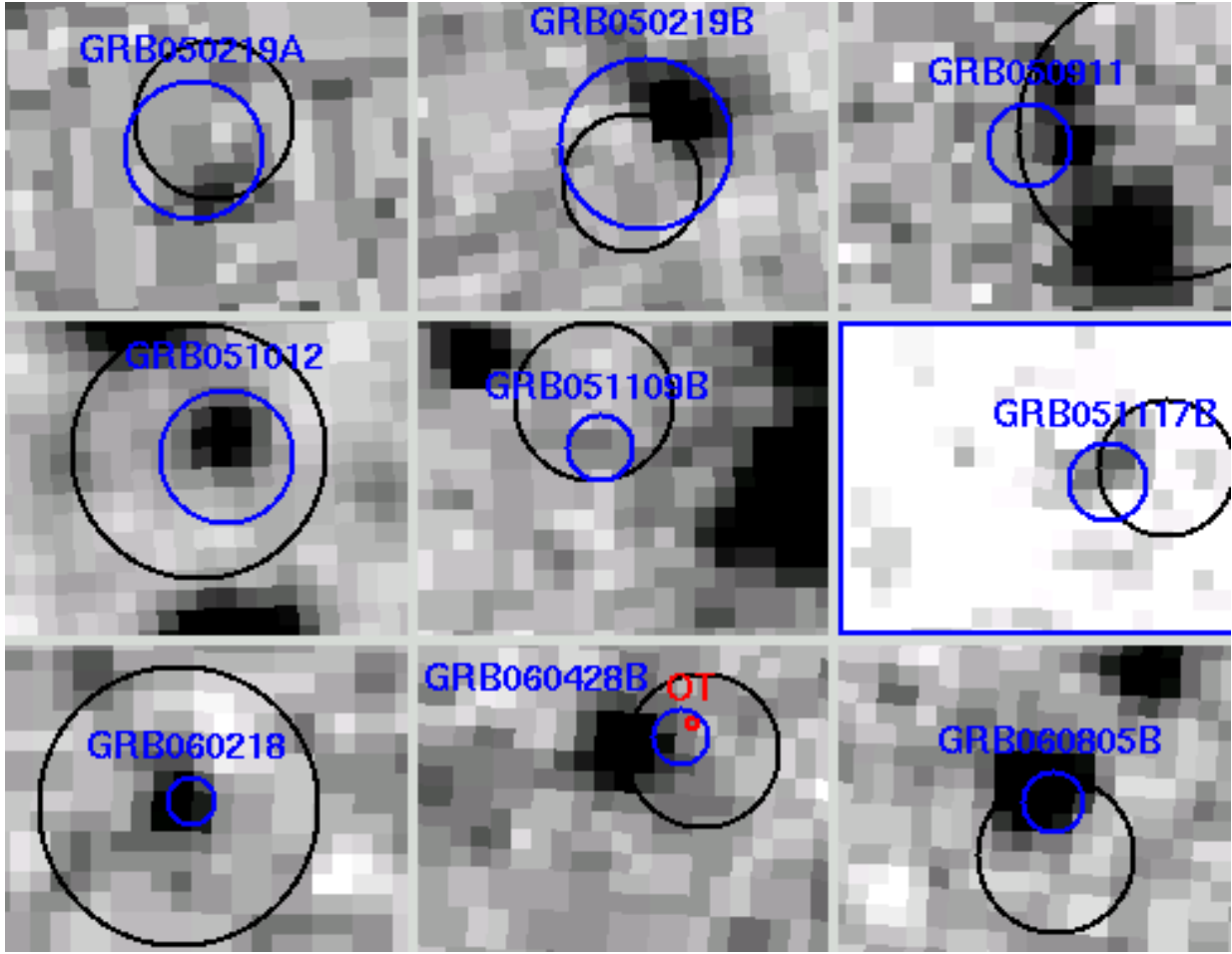}}
\caption{\small DSS images of nine refined X-ray positions regions (circles) which are consistent with bright optical sources.
Aside from the bright spiral galaxy to the West of GRB~051109B (with $R\sim 14.5$), the
optical magnitudes range from $R=18.3-20.5$.  Our error regions are shown in blue, with the XRT team error circles in black.
In most cases (but not all) the refined error regions are smaller.  In all cases, the refined error regions better identify 
the optical association.}
\label{fig:f5}
\end{figure}

The refined positions may be most useful for host galaxy studies.
For example, there is evidence for an over-density of bright optical sources within our error regions.
The optical/X-ray match fractions translate to an average optical
source density of $(1.55\pm 0.03) \times 10^{-3}$ arcsec$^{-2}$ ($R\lessim 20$) and $(2.74\pm 0.09) \times 10^{-3}$ arcsec$^{-2}$ ($r'\lessim 22.5$).
There are enough (120) DSS images with error radii $<5\arcsec$~to allow for the determination of the
density within the error region: 8 sources or a density of $(4\pm1) \times 10^{-3}$ arcsec$^{-2}$ (Figure \ref{fig:f5}).  We conclude from this
significant over-density that most of these sources---aside from one associated with GRB~050721 which occurred at Galactic latitude
$|b|<10$---may be bright GRB host galaxies.  Consonant with an interpretation that these events are relatively nearby,
$\sim 3$ are X-ray Flashes.  We note that none are short/hard GRBs.  
The only burst with a measured redshift is GRB/XRF~060218 \citep[$z=0.0331$;][]{mirabaletal06}.
The SDSS image shows a spatially extended host galaxy \citep{mirabal06}.  
Several of the other possible host galaxies in Figure \ref{fig:f5} appear to possibly be extended.
The only other source in the list of 8 which has a detected OT is GRB/XRF~060428B.  
\citet{perley06b} have explored the possibility that this nearby ($z=0.348$) galaxy may physically associated and conclude  
that it is not.
Finally, although the core of the potential host galaxy is not in our error region,  \citet{perley06a} report 
that our position for GRB~051109B lies in the spiral arm of a bright galaxy (Figure \ref{fig:f5}).

\section{Conclusions}

We present a catalog of 179 positions for the X-ray afterglows of {\it Swift}~GRBs, with astrometry solutions
determined by matching optical positions to those of X-ray field sources.  On average, the position are determined
with a factor of two more accuracy and with 50\% smaller error radius estimates than when the astrometry correction is
neglected and the satellite aspect solution is assumed.  The typical position error is 2.2\arcsec, with no
systematic error term.  There is now very little chance ($<2$\%) of source confusion in the XRT error regions.  The
properties of the optical/X-ray field sources observed in 1.9 Msec of PC mode data appear to be similar to
sources observed (in the bright end) in deep field surveys.  There is an indication
of bright host galaxies in our error regions of nine afterglows.  We hope that our positions will facilitate
deep observations of potential host galaxies for a large fraction of past and future GRBs with {\it Swift}~XRT afterglows.

\acknowledgments
N. Butler gratefully acknowledges support from a Townes Fellowship at the U.~C. Berkeley 
Space Sciences Laboratory, as well as partial support from J. Bloom and A. Filippenko.  Special
thanks to J. Bloom for many useful discussions.  We thank an anonymous referee for useful and
insightful criticism.

\clearpage

\input{tab1.tex}
\input{tab2.tex}
\input{tab3.tex}
\end{document}

%% file: tab1.tex
\begin{table}
\begin{center}
\caption{Astrometrically-corrected XRT GRB Afterglow Positions}
\vspace{1mm}
\footnotesize
\begin{tabular}{lccclccc}\hline\hline
GRB & $\alpha$ (J2000) & $\delta$ (J2000) & Err$_{90}$ [\arcsec] & 
GRB & $\alpha$ (J2000) & $\delta$ (J2000) & Err$_{90}$ [\arcsec] \\\hline
050408$_{16}$ &  12 02 17.36  &  +10 51 10.5  & 1.2$^{a}$ & 	050505$_{4}$ &  09 27 03.34  &  +30 16 24.6  & 0.8$^{a}$ \\ 
050504 &  13 24 01.18  &  +40 42 14.3  & 4.6$^{a}$ & 	050509A &  20 42 19.88  &  +54 04 19.4  & 2.5$^{b}$ \\ 
050520 & \underline{  12 50 05.94 } & \underline{  +30 27 03.2 } & 2.4$^{a}$ & 	050509B &  12 36 14.06  &  +28 59 07.2  & 3.4$^{a}$ \\ 
050522 &  13 20 34.57  &  +24 47 20.7  & 3.8$^{a}$ & 	050525$_{48}$ &  18 32 32.67  &  +26 20 21.2  & 2.1$^{b}$ \\ 
050714A &  02 54 23.33  &  +69 06 45.1  & 2.7$^{b}$ & 	050528 & \underline{  23 34 04.35 } & \underline{  +45 58 21.8 } & 2.9$^{b}$ \\ 
051012 &  18 02 10.95  &  -52 47 12.6  & 3.4$^{b}$ & 	050603$_{7}$ &  02 39 56.90  &  -25 10 55.7  & 0.9$^{b}$ \\ 
051021A$_{47}$ &  01 56 36.41  &  +09 04 04.3  & 2.1$^{b}$ & 	050607$_{29}$ &  20 00 42.77  &  +09 08 31.1  & 1.4$^{b}$ \\ 
051022$_{12}$ &  23 56 04.08  &  +19 36 23.3  & 1.1$^{b}$ & 	050701 &  15 09 01.67  &  -59 24 55.6  & 2.4$^{b}$ \\ 
051028$_{59}$ & \underline{  01 48 14.97 } & \underline{  +47 45 08.3 } & 2.8$^{b}$ & 	050712 &  05 10 48.34  &  +64 54 46.9  & 1.4$^{b}$ \\ 
051105B &  00 37 53.73  &  -40 29 16.3  & 2.8$^{b}$ & 	050713A$_{61}$ &  21 22 09.54  &  +77 04 29.7  & 2.8$^{b}$ \\ 
051211B$_{35}$ &  23 02 41.50  &  +55 04 50.8  & 1.6$^{b}$ & 	050713B &  20 31 15.59  &  +60 56 41.9  & 1.0$^{b}$ \\ 
060121$_{22}$ &  09 09 52.00  &  +45 39 45.4  & 1.4$^{a}$ & 	050714B &  11 18 47.75  &  -15 32 49.3  & 2.1$^{b}$ \\ 
060123 &  11 58 47.83  &  +45 30 50.9  & 1.0$^{a}$ & 	050716 &  22 34 20.72  &  +38 41 03.6  & 0.9$^{b}$ \\ 
060805B &  18 10 05.57  &  +58 09 18.9  & 1.5$^{b}$ & 	050717 &  14 17 24.61  &  -50 31 59.9  & 3.6$^{b}$ \\ 
060901 &  19 08 38.02  &  -06 38 09.0  & 6.7$^{b}$ & 	050721$_{67}$ &  16 53 44.40  &  -28 22 52.2  & 3.4$^{b}$ \\ 
060928 & \underline{  08 30 30.53 } & \underline{  -42 45 01.8 } & 10.2$^{b}$ & 	050724$_{71}$ & \underline{  16 24 44.32 } & \underline{  -27 32 26.4 } & 5.0$^{b}$ \\ 
061025 & \underline{  20 03 37.18 } & \underline{  -48 14 28.9 } & 3.4$^{b}$ & 	050726 &  13 20 12.16  &  -32 03 51.0  & 3.7$^{b}$ \\ 
041223 &  06 40 47.39  &  -37 04 22.6  & 6.4$^{b}$ & 	050730$_{28}$ &  14 08 17.12  &  -03 46 16.3  & 1.4$^{b}$ \\ 
050117 & \underline{  23 53 47.24 } & \underline{  +65 56 00.2 } & 14.1$^{b}$ & 	050801$_{72}$ &  13 36 35.51  &  -21 55 42.7  & 5.0$^{b}$ \\ 
050124 &  12 51 30.43  &  +13 02 40.6  & 2.8$^{b}$ & 	050802$_{33}$ &  14 37 05.91  &  +27 47 11.3  & 1.5$^{a}$ \\ 
050126 &  18 32 27.22  &  +42 22 14.2  & 3.3$^{b}$ & 	050803 &  23 22 37.87  &  +05 47 09.8  & 1.8$^{b}$ \\ 
050128 &  14 38 17.66  &  -34 45 54.7  & 2.7$^{b}$ & 	050813 & \underline{  16 07 57.19 } & \underline{  +11 14 57.8 } & 3.8$^{a}$ \\ 
050215B$_{44}$ &  11 37 47.74  &  +40 47 44.6  & 1.9$^{a}$ & 	050814$_{10}$ &  17 36 45.31  &  +46 20 21.6  & 1.0$^{b}$ \\ 
050219A &  11 05 39.13  &  -40 41 02.6  & 3.5$^{b}$ & 	050815$_{76}$ & \underline{  19 34 22.80 } & \underline{  +09 08 47.5 } & 7.9$^{b}$ \\ 
050219B &  05 25 15.76  &  -57 45 28.2  & 4.3$^{b}$ & 	050819 & \underline{  23 55 01.49 } & \underline{  +24 51 39.9 } & 3.3$^{b}$ \\ 
050223 &  18 05 32.66  &  -62 28 20.4  & 3.2$^{b}$ & 	050820A$_{6}$ &  22 29 38.08  &  +19 33 36.4  & 0.9$^{b}$ \\ 
050306 &  18 49 14.43  &  -09 09 09.9  & 4.7$^{b}$ & 	050820B &  09 02 25.37  &  -72 38 43.7  & 5.2$^{b}$ \\ 
050315$_{24}$ &  20 25 54.21  &  -42 36 02.5  & 1.4$^{b}$ & 	050822 &  03 24 27.22  &  -46 02 00.0  & 0.7$^{b}$ \\ 
050318$_{58}$ &  03 18 51.04  &  -46 23 43.5  & 2.7$^{b}$ & 	050824 &  00 48 56.16  &  +22 36 33.4  & 1.0$^{b}$ \\ 
050319$_{13}$ &  10 16 47.89  &  +43 32 53.3  & 1.1$^{a}$ & 	050826$_{55}$ &  05 51 01.69  &  -02 38 37.6  & 2.6$^{b}$ \\ 
050326 &  00 27 49.11  &  -71 22 16.4  & 2.9$^{b}$ & 	050827 &  04 17 09.58  &  +18 12 00.4  & 1.3$^{b}$ \\ 
050401$_{41}$ &  16 31 28.84  &  +02 11 14.5  & 1.8$^{b}$ & 	050904 & \underline{  00 54 50.93 } & \underline{  +14 05 11.8 } & 1.2$^{a}$ \\ 
050406$_{19}$ &  02 17 52.25  &  -50 11 15.0  & 1.3$^{b}$ & 	050906 &  03 31 15.28  &  -14 36 30.1  & 15.7$^{b}$ \\ 
050410 & \underline{  05 59 13.94 } & \underline{  +79 36 11.7 } & 2.3$^{b}$ & 	050908$_{49}$ &  01 21 50.85  &  -12 57 17.9  & 2.2$^{b}$ \\ 
050412$_{73}$ &  12 04 25.18  &  -01 12 00.8  & 6.9$^{a}$ & 	050911 &  00 54 38.29  &  -38 50 58.5  & 2.1$^{b}$ \\ 
050416A$_{2}$ &  12 33 54.57  &  +21 03 26.9  & 0.6$^{a}$ & 	050915A$_{26}$ &  05 26 44.86  &  -28 00 59.9  & 1.4$^{b}$ \\ 
050416B &  08 55 34.91  &  +11 10 30.7  & 1.8$^{b}$ & 	050915B &  14 36 26.24  &  -67 24 31.7  & 4.1$^{b}$ \\ 
050421 &  20 29 03.22  &  +73 39 18.9  & 3.3$^{b}$ & 	050916 &  09 03 57.07  &  -51 25 43.4  & 5.7$^{b}$ \\ 
050422 &  21 37 54.65  &  +55 46 46.4  & 1.5$^{b}$ & 	050922B &  00 23 13.19  &  -05 36 15.6  & 2.7$^{b}$ \\ 
050502B &  09 30 10.06  &  +16 59 46.5  & 1.0$^{a}$ & 	050922C$_{46}$ &  21 09 33.12  &  -08 45 28.3  & 2.0$^{b}$ \\ 
\hline
\end{tabular}
\end{center}
{\small Notes: Positions lying outside the refined XRT-Team error regions are underlined. $^a$- Position matching with SDSS DR-5 catalog. $^b$- with DSS-red2 catalog.
The subscripts on some GRB names refer to Figure \ref{fig:f3} and denote the bursts with OT positions.  The first 17 bursts above were discovered
by satellites other than Swift.}
\label{tab:pos1}
\end{table}

%% file: tab2.tex
\begin{table}
\begin{center}
\caption{Astrometrically-corrected XRT GRB Afterglow Positions (continued)}
\vspace{1mm}
\footnotesize
\begin{tabular}{lccclccc}\hline\hline
GRB & $\alpha$ (J2000) & $\delta$ (J2000) & Err$_{90}$ [\arcsec] &
GRB & $\alpha$ (J2000) & $\delta$ (J2000) & Err$_{90}$ [\arcsec] \\\hline
050925 &  20 13 47.57  &  +34 19 51.4  & 2.7$^{b}$ & 	060322 &  18 16 56.54  &  -36 42 32.4  & 6.7$^{b}$ \\ 
051001 &  23 23 48.73  &  -31 31 23.3  & 1.5$^{b}$ & 	060323 &  11 37 45.19  &  +49 59 06.4  & 2.0$^{a}$ \\ 
051006 &  07 23 14.03  &  +09 30 21.9  & 4.3$^{b}$ & 	060403 &  18 49 21.44  &  +08 19 44.6  & 3.1$^{b}$ \\ 
051008 &  13 31 29.55  &  +42 05 53.3  & 1.2$^{a}$ & 	060413 & \underline{  19 25 07.91 } & \underline{  +13 45 30.1 } & 1.9$^{b}$ \\ 
051016A &  08 11 16.77  &  -18 17 53.7  & 2.2$^{b}$ & 	060418$_{70}$ &  15 45 42.47  &  -03 38 20.1  & 4.9$^{b}$ \\ 
051016B$_{8}$ &  08 48 27.80  &  +13 39 20.7  & 0.9$^{a}$ & 	060421 &  22 54 32.87  &  +62 43 50.1  & 1.6$^{b}$ \\ 
051021B &  08 24 11.98  &  -45 32 28.2  & 3.4$^{b}$ & 	060427 &  08 17 04.53  &  +62 40 19.5  & 7.2$^{b}$ \\ 
051109A$_{15}$ &  22 01 15.32  &  +40 49 21.6  & 1.1$^{b}$ & 	060428A &  08 14 10.88  &  -37 10 11.2  & 0.8$^{b}$ \\ 
051109B &  23 01 50.32  &  +38 40 47.3  & 1.6$^{b}$ & 	060428B$_{27}$ &  15 41 25.72  &  +62 01 29.6  & 1.4$^{a}$ \\ 
051111 &  23 12 33.22  &  +18 22 28.5  & 2.6$^{b}$ & 	060501 & \underline{  21 53 30.46 } & \underline{  +43 59 54.8 } & 8.8$^{b}$ \\ 
051117A$_{31}$ & \underline{  15 13 34.00 } & \underline{  +30 52 10.9 } & 1.5$^{a}$ & 	060502A$_{45}$ & \underline{  16 03 42.56 } & \underline{  +66 36 02.9 } & 2.0$^{b}$ \\ 
051117B &  05 40 43.21  &  -19 16 27.2  & 2.0$^{b}$ & 	060502B$_{68}$ &  18 35 45.53  &  +52 37 52.9  & 3.7$^{b}$ \\ 
051210 &  22 00 41.26  &  -57 36 46.5  & 2.9$^{b}$ & 	060505 & \underline{  22 07 03.39 } & \underline{  -27 48 53.0 } & 2.5$^{b}$ \\ 
051221A$_{9}$ &  21 54 48.69  &  +16 53 28.0  & 0.9$^{b}$ & 	060507 & \underline{  05 59 50.62 } & \underline{  +75 14 57.2 } & 1.4$^{b}$ \\ 
051221B &  20 49 35.10  &  +53 02 12.9  & 6.9$^{b}$ & 	060510A$_{42}$ &  06 23 27.90  &  -01 09 45.5  & 1.9$^{b}$ \\ 
051227$_{57}$ & \underline{  08 20 58.06 } & \underline{  +31 55 34.2 } & 2.7$^{a}$ & 	060510B &  15 56 30.07  &  +78 34 12.4  & 2.9$^{b}$ \\ 
060105 &  19 50 00.69  &  +46 20 55.9  & 1.3$^{b}$ & 	060512$_{50}$ &  13 03 05.73  &  +41 11 26.9  & 2.4$^{a}$ \\ 
060108$_{21}$ & \underline{  09 48 02.00 } & \underline{  +31 55 08.0 } & 1.3$^{a}$ & 	060515 &  08 29 09.49  &  +73 34 00.3  & 4.7$^{b}$ \\ 
060109 &  18 50 43.55  &  +31 59 28.3  & 1.8$^{b}$ & 	060522$_{62}$ & \underline{  21 31 45.02 } & \underline{  +02 53 13.0 } & 2.8$^{b}$ \\ 
060110$_{63}$ &  04 50 57.76  &  +28 25 55.0  & 2.9$^{b}$ & 	060526$_{34}$ &  15 31 18.23  &  +00 17 05.4  & 1.6$^{a}$ \\ 
060111A &  18 24 49.20  &  +37 36 14.1  & 1.1$^{b}$ & 	060602A &  09 58 16.93  &  +00 18 12.8  & 2.3$^{b}$ \\ 
060111B$_{54}$ & \underline{  19 05 42.75 } & \underline{  +70 22 33.3 } & 2.6$^{b}$ & 	060602B & \underline{  17 49 31.94 } & \underline{  -28 08 05.8 } & 3.3$^{b}$ \\ 
060115 & \underline{  03 36 08.18 } & \underline{  +17 20 44.3 } & 2.6$^{b}$ & 	060604$_{11}$ & \underline{  22 28 55.01 } & \underline{  -10 54 55.9 } & 1.0$^{b}$ \\ 
060116$_{64}$ &  05 38 46.14  &  -05 26 15.2  & 3.1$^{b}$ & 	060605$_{60}$ &  21 28 37.30  &  -06 03 32.2  & 2.8$^{b}$ \\ 
060124 &  05 08 26.00  &  +69 44 27.2  & 0.7$^{b}$ & 	060607A$_{37}$ & \underline{  21 58 50.44 } & \underline{  -22 29 48.1 } & 1.7$^{b}$ \\ 
060202 &  02 23 23.01  &  +38 23 03.2  & 1.1$^{b}$ & 	060614$_{23}$ &  21 23 32.04  &  -53 01 37.0  & 1.4$^{b}$ \\ 
060203$_{14}$ &  06 54 04.05  &  +71 48 39.3  & 1.1$^{b}$ & 	060707$_{36}$ &  23 48 19.07  &  -17 54 18.9  & 1.6$^{b}$ \\ 
060204B & \underline{  14 07 15.05 } & \underline{  +27 40 36.9 } & 1.6$^{a}$ & 	060708$_{25}$ &  00 31 13.79  &  -33 45 33.3  & 1.4$^{b}$ \\ 
060206$_{3}$ &  13 31 43.51  &  +35 03 02.8  & 0.8$^{a}$ & 	060712 &  12 16 16.12  &  +35 32 17.7  & 1.5$^{a}$ \\ 
060210$_{5}$ &  03 50 57.35  &  +27 01 34.3  & 0.9$^{b}$ & 	060714$_{39}$ &  15 11 26.43  &  -06 33 59.5  & 1.8$^{b}$ \\ 
060211A &  03 53 32.63  &  +21 29 19.4  & 2.0$^{b}$ & 	060717 &  11 23 21.74  &  +28 57 04.9  & 4.3$^{a}$ \\ 
060211B &  05 00 17.12  &  +14 56 58.1  & 1.9$^{b}$ & 	060719$_{30}$ & \underline{  01 13 43.69 } & \underline{  -48 22 51.0 } & 1.5$^{b}$ \\ 
060218$_{18}$ &  03 21 39.66  &  +16 52 02.1  & 1.2$^{b}$ & 	060729$_{1}$ & \underline{  06 21 31.90 } & \underline{  -62 22 12.6 } & 0.5$^{b}$ \\ 
060219 & \underline{  16 07 21.54 } & \underline{  +32 18 57.3 } & 1.6$^{a}$ & 	060801 & \underline{  14 12 01.35 } & \underline{  +16 58 53.7 } & 2.4$^{a}$ \\ 
060223A$_{66}$ &  03 40 49.82  &  -17 07 49.8  & 3.4$^{b}$ & 	060804$_{77}$ & \underline{  07 28 49.60 } & \underline{  -27 12 52.8 } & 10.8$^{b}$ \\ 
060223B & \underline{  16 56 58.67 } & \underline{  -30 48 35.4 } & 12.4$^{b}$ & 	060805A &  14 43 43.46  &  +12 35 11.8  & 1.8$^{a}$ \\ 
060306 &  02 44 22.91  &  -02 08 54.0  & 1.3$^{b}$ & 	060807$_{17}$ & \underline{  16 50 02.58 } & \underline{  +31 35 30.4 } & 1.2$^{a}$ \\ 
060312 &  03 03 05.92  &  +12 50 02.0  & 2.0$^{b}$ & 	060813 & \underline{  07 27 35.23 } & \underline{  -29 50 50.3 } & 2.2$^{b}$ \\ 
060313$_{52}$ &  04 26 28.41  &  -10 50 40.7  & 2.4$^{b}$ & 	060814 &  14 45 21.29  &  +20 35 10.7  & 0.9$^{a}$ \\ 
060319 & \underline{  11 45 32.89 } & \underline{  +60 00 39.1 } & 0.9$^{b}$ & 	060825 & \underline{  01 12 29.46 } & \underline{  +55 47 51.4 } & 2.9$^{b}$ \\ 
\hline
\end{tabular}
\end{center}
{\small Notes: Positions lying outside the refined XRT-Team error regions are underlined.  $^a$- Position matching with SDSS DR-5 catalog. $^b$- with DSS-red2 catalog.
The subscripts on some GRB names refer to Figure \ref{fig:f3} and denote the bursts with OT positions.}
\label{tab:pos2}
\end{table}

%% file: tab3.tex
\begin{table}
\begin{center}
\caption{Astrometrically-corrected XRT GRB Afterglow Positions (continued)}
\vspace{1mm}
\footnotesize
\begin{tabular}{lccclccc}\hline\hline
GRB & $\alpha$ (J2000) & $\delta$ (J2000) & Err$_{90}$ [\arcsec] &
GRB & $\alpha$ (J2000) & $\delta$ (J2000) & Err$_{90}$ [\arcsec] \\\hline
060904A &  15 50 54.56  &  +44 59 10.5  & 1.4$^{b}$ & 	060927 &  21 58 12.23  &  +05 21 52.2  & 3.8$^{b}$ \\ 
060904B$_{38}$ & \underline{  03 52 50.45 } & \underline{  -00 43 30.0 } & 1.7$^{a}$ & 	060929 &  17 32 28.98  &  +29 50 07.7  & 2.2$^{b}$ \\ 
060906$_{53}$ & \underline{  02 43 00.72 } & \underline{  +30 21 43.2 } & 2.5$^{b}$ & 	061002 &  14 41 23.40  &  +48 44 28.9  & 2.2$^{b}$ \\ 
060908$_{56}$ & \underline{  02 07 18.42 } & \underline{  +00 20 30.8 } & 2.6$^{b}$ & 	061004 &  06 31 10.93  &  -45 54 23.8  & 1.6$^{b}$ \\ 
060912A$_{43}$ &  00 21 08.11  &  +20 58 18.9  & 1.9$^{b}$ & 	061006$_{51}$ &  07 24 07.56  &  -79 11 56.6  & 2.4$^{b}$ \\ 
060919 &  18 27 41.89  &  -51 00 51.1  & 2.6$^{b}$ & 	061007$_{32}$ &  03 05 19.55  &  -50 30 01.8  & 1.5$^{b}$ \\ 
060923A$_{65}$ &  16 58 28.05  &  +12 21 39.5  & 3.4$^{b}$ & 	061019$_{74}$ &  06 06 30.80  &  +29 34 11.0  & 6.9$^{b}$ \\ 
060923B &  15 52 46.74  &  -30 54 11.3  & 7.4$^{b}$ & 	061021$_{20}$ &  09 40 36.04  &  -21 57 04.9  & 1.3$^{b}$ \\ 
060923C$_{40}$ &  23 04 28.36  &  +03 55 28.4  & 1.8$^{b}$ & 	061028 &  06 28 54.61  &  +46 17 57.5  & 2.7$^{b}$ \\ 
060926$_{69}$ &  17 35 43.86  &  +13 02 19.0  & 4.6$^{b}$ & 	& & & \\
\hline
\end{tabular}
\end{center}
{\small Notes: Positions lying outside the refined XRT-Team error regions are underlined.  $^a$- Position matching with SDSS DR-5 catalog. $^b$- with DSS-red2 catalog.
The subscripts on some GRB names refer to Figure \ref{fig:f3} and denote the bursts with OT positions.}
\label{tab:pos3}
\end{table}